\documentstyle[12pt]{article}
\begin{document}
\title{ $\sigma$ Decay at Finite Temperature and Density }
\author{P. Zhuang and Z. Yang\\
        Physics Department, Tsinghua University, Beijing 100084, China}
\date{}
\maketitle

\begin{abstract}
$\sigma$ decay and its relation with chiral phase transition are discussed at finite temperature and density in the framework of the Nambu-Jona-Lasinio model. The decay rate for the process $\sigma\rightarrow 2\pi$ to first order in a $1/N_c$ expansion is calculated as a function of temperature $T$ and baryon density $n_b$. In particular, only when the chiral phase transition happens around the tricritical point, the $\sigma$ decay results in a non-thermal enhancement of pions in the final state distributions in relativistic heavy ion collisions.  
\end{abstract}

It is generally believed that there are two QCD phase transitions in hot and dense nuclear matter\cite{MuHe}. One of them is related to the deconfinement process in moving from a hadron gas to a quark-gluon plasma, and the other one describes the transition from the chiral symmetry breaking phase to the phase in which it is restored. 

At low energies, there is no obvious reason for the presence of the scalar meson $\sigma$ in the study of chiral properties. However, there is a definite need\cite{Rho} for $\sigma$ at finite temperature and density in the investigation of chiral symmetry restoration. For instance\cite{Zhuang}, the contribution from $\sigma$ to thermodynamics can for most case be neglected, because of its heavy mass, while  it plays an important role in the region around the critical point of the chiral phase transition, since there, $m_\sigma=0$. Because it is still difficult to extract definite information from the lattice simulations with nonzero baryon density, we need QCD models to investigate the phase transitions at finite density. From the study of several models\cite{Rapp}, the chiral phase transition in the chiral limit with massless pions is of second order at high temperature $T$, and of first order at high density $n_b$. Therefore, there is a tricritical point $P$ in the $T-n_b$ plane which separates the first- and second-order phase transitions. In the real world with nonzero pion mass, the second-order phase transition becomes a smooth crossover and the tricritical point $P$ becomes a critical end-point $E$ of a first-order phase transition. In relativistic heavy ion collisions, for the choice of control parameters like colliding energy and impact parameter such that the freeze-out of the system occurs near the point $E$, $\sigma$ is one of the most numerous species at the freeze-out, since it is nearly massless. During the expansion of the system, the in-medium $\sigma$ mass rises towards its vacuum value and eventually exceeds the $\pi\pi$ threshold. As the $\sigma\pi\pi$ coupling is large, the decay proceeds rapidly. Since this process occurs after freeze-out, the pions generated by it do not have a chance to thermalize. Thus, it is expected\cite{Rajagopal} that the resulting pion spectrum should have a non-thermal enhancement at low transverse momentum. 

In this letter we discuss the decay process $\sigma\rightarrow 2\pi$ at finite temperature and density in the framework of the Nambu-Jona-lasinio (NJL) model\cite{NJL} which is believed to describe the chiral properties well. This was firstly studied at zero density by Hatsuda and Kunihiro\cite{NJL}. Our motivation here is to see that in which temperature and density region of freeze-out there is a non-thermal enhancement of pions due to the $\sigma$ decay, by calculating the decay rate in the whole $T-n_b$ plane. 

The two-flavor version of the NJL model is defined through the lagrangian density,
\begin{equation}
\label{njl}
L_{NJL} = \bar\psi (i\gamma^\mu\partial_\mu - m_0)\psi+G[(\bar\psi\psi)^2+(\bar\psi i \gamma_5\tau\psi)^2],
\end{equation}
where only the scalar and pseudoscalar interactions corresponding to $\sigma$ and $\pi$ mesons, respectively, are considered, $\psi$ and $\bar\psi$ are the quark fields, $\tau$ is the $SU(2)$ isospin generator, $G$ is the coupling constant with dimension $GeV^{-2}$, and $m_0$ is the current quark mass.

In an expansion in the inverse number of colors, $1/N_c$, the zeroth order (Hartree) approximation for quarks together with the first order (RPA) approximation for mesons gives a self-consistent treatment of the quark-meson plasma in the NJL model. To the lowest order, namely $\left(1/N_c\right)^{1/2}$, the Feynman diagram for the decay process $\sigma\rightarrow 2\pi$ is sketched in Fig.1. The partial decay rate in the rest frame of $\sigma$ is given in terms of the Lorentz-invariant matrix element $M$ by 
\begin{equation}
\label{rate1}
{d\Gamma_{\sigma\rightarrow 2\pi}\over d\Omega} = {1\over 32\pi^2}{|{\bf p}|\over m_\sigma^2}|M|^2\ ,
\end{equation}
where $|{\bf p}|=\sqrt{{m_\sigma^2\over 4}-m_\pi^2}$ is the pion momentum, $m_\sigma$ and $m_\pi$ are respectively $\sigma$ and $\pi$ masses. using an obvious notation for the $M$-matrix element, one has
\begin{equation}
\label{matrix}
M=g_\sigma g_\pi^2 A_{\sigma\pi\pi}
\end{equation}
with
\begin{equation}
\label{aspp}
A_{\sigma\pi\pi} = T\sum_n e^{i\omega_n\eta}\int{d^3{\bf q}\over (2\pi)^3}Tr S({\bf q},i\omega_n)\Gamma_\pi S({\bf q}+{\bf p},i\omega_n+{m_\sigma\over 2})\Gamma_\pi S({\bf q},i\omega_n +m_\sigma)\ .
\end{equation}
Here $\omega_n = (2n+1)\pi T, n=0,\pm 1,\pm 2,...$ are fermionic Matsubara frequencies, $\Gamma_\pi = i\gamma_5\tau$ is the pseudoscalar interaction vertex. The quark propagator is denoted by $S({\bf q},i\omega_n) = \left(\gamma_0(i\omega_n)-{\bf \gamma}\cdot{\bf q}+m_q\right)/\left((i\omega_n)^2-E_q^2\right)$, with $E_q^2 = m_q^2+{\bf q}^2$. In Eq.(\ref{aspp}), $Tr$ refers to the trace over color, flavor and spinor indices. The pion-quark and sigma-quark coupling strengths in Eq.(\ref{matrix}) are determined in the model via\cite{NJL}
\begin{eqnarray}
\label{strength}
&& g_\pi^{-2}(T,\mu) = {\partial \Pi_\pi(k_0,{\bf 0};T,\mu)\over \partial k_0^2}|_{k_0^2=m_\pi^2}\ ,\nonumber\\
&& g_\sigma^{-2}(T,\mu) = {\partial \Pi_\sigma(k_0,{\bf 0};T,\mu)\over \partial k_0^2}|_{k_0^2=m_\sigma^2}\ ,
\end{eqnarray}
where $\Pi_\pi$ and $\Pi_\sigma$ are the standard mesonic polarization functions\cite{NJL} for $\pi$ and $\sigma$. The dynamically generated quark mass $m_q(T,\mu)$ and the meson masses $m_\pi(T,\mu)$ and $m_\sigma(T,\mu)$ are calculated using the usual gap equations\cite{NJL} in the Hartree approximation for quarks and RPA approximation for mesons. After evaluation of the Matsubara sum in Eq.(\ref{aspp}), one has\cite{Huf}
\begin{eqnarray}
\label{aspp1}
A_{\sigma\pi\pi}(T,\mu) = && 4m_q N_c N_f\int{d^3{\bf q}\over (2\pi)^3}{f_F(E_q-\mu)-f_F(-E_q-\mu)\over 2E_q}\times \nonumber\\
&& {8({\bf q}\cdot{\bf p})^2-(2m_\sigma^2+4m_\pi^2){\bf q}\cdot{\bf p}+m_\sigma^2/2-2m_\sigma^2 E_q^2\over (m_\sigma^2-4E_q^2)\left((m_\pi^2-2{\bf q}\cdot{\bf p})^2-m_\sigma^2 E_q^2\right)}\ ,
\end{eqnarray} 
where $f_F(x) = \left(1+e^{x/T}\right)^{-1}$ is the Fermi-Dirac distribution function.

$\sigma$ can decay into neutral and charged pions. By considering the exchange contribution and $0\le \theta \le \pi/2$ for $\sigma\rightarrow 2\pi_0$ due to the identical particle effect, the total decay rate can be written as
\begin{eqnarray}
\label{rate2}
\Gamma_{\sigma\rightarrow 2\pi}(T,\mu) && = \Gamma_{\sigma\rightarrow 2\pi_0}(T,\mu)+\Gamma_{\sigma\rightarrow \pi_+\pi_-}(T,\mu)\nonumber\\
&& = {3\over 8\pi}{\sqrt{{m_\sigma^2\over 4}-m_\pi^2}\over m_\sigma^2}g_\sigma^2 g_\pi^4 |A_{\sigma\pi\pi}(T,\mu)|^2\left(1+f_B({m_\sigma\over 2})\right)^2\ ,
\end{eqnarray}
where we have taken into account the Bose-Einstein statistics in terms of the distribution function $f_B(x) = \left(e^{x/T}-1\right)^{-1}$ for the final state pions. 
    
From the comparison of $\sigma-\pi-\pi$ coupling strength $g_{\sigma\pi\pi}$ defined through the lagrangian $L_I \sim g_{\sigma\pi\pi}\sigma{\bf \pi}\cdot{\bf \pi}$ and the $M$-matrix element (\ref{matrix}) in the NJL model, we have
\begin{equation}
\label{gspp}
g_{\sigma\pi\pi}(T,\mu) = 2M(T,\mu) = 2g_\sigma g_\pi^2 A(T,\mu)\ .
\end{equation}

Now let us turn to numerical calculations of the coupling strength $g_{\sigma\pi\pi}$ and the decay rate $\Gamma_{\sigma\rightarrow 2\pi}$ as functions of temperature $T$ and baryon density $n_b$. The relation between $n_b$ and the chemical potential $\mu$ is rather complicated and is also a function of $T$,
\begin{equation}
\label{nb}
n_b(T,\mu) = {N_c N_f\over 3}\int{d^3{\bf q}\over (2\pi)^3}\left[\tanh{1\over 2T}(E_q+\mu)-\tanh{1\over 2T}(E_q-\mu)\right]\ ,
\end{equation}
where the factor $3$ reflects the fact that $3$ quarks make a baryon. In the calculation we use the dynamical quark mass $m_q = 0.32\ GeV$, the pion decay constant $f_\pi = 0.093\ GeV$ and the pion mass $m_\pi = 0.134\ GeV$ in the vacuum ($T=\mu=0$) as input to determine the $3$ parameters in the NJL model, namely the coupling constant $G=4.93\ GeV^{-2}$, the momentum cutoff $\Lambda=0.653\ GeV$ and the current quark mass $m_0 = 0.005\ GeV$.

We first discuss the temperature dependence of $g_{\sigma\pi\pi}$ and $\Gamma_{\sigma\rightarrow 2\pi}$ in the case of zero baryon density which corresponds to the central region of relativistic heavy ion collisions. The coupling strength $g_{\sigma\pi\pi}$ is about $2\ GeV$ in the vacuum and almost a constant in the temperature region $T< 0.1\ GeV$, and then decreases smoothly with increase of the temperature, as shown in Fig.2. The energy condition for the decay process $\sigma\rightarrow 2\pi$ to happen is $m_\sigma > 2m_\pi$ in the rest frame of $\sigma$. At the threshold temperature $T\simeq 0.19\ GeV$ determined by $ m_\sigma(T)=2m_\pi(T)$, $g_{\sigma\pi\pi}$ has a sudden jump from about $1.2\ GeV$ to zero. In a wide temperature region, the decay rate (\ref{rate2}) goes up gradually from its vacuum value $\sim 0.09\ GeV$ to the maximum $\sim 0.11\ GeV$ at $T\simeq 0.17\ GeV$, and then it drops down rapidlly but continuously to zero at the threshold temperature. The fact that the maximum value of $\Gamma$ is not in the vacuum, but at $T\simeq 0.17\ GeV$ is beyond our expectation. It originates from the important Bose-Einstein statistics of the final state pions at high temperature. This can be seen clearly in Fig.2 where we show the decay rates with and without consideration of the Bose-Einstein statistics. The rapid decrease of $\Gamma_{\sigma\rightarrow 2\pi}$ at high temperature is due to the behavior of $g_{\sigma\pi\pi}$ and the phase-space factor $\sqrt{m_\sigma^2/4-m_\pi^2}/m_\sigma^2$.    

The density effect on $g_{\sigma\pi\pi}$ and $\Gamma_{\sigma\rightarrow 2\pi}$ is shown in Fig.3. Different from the temperature effect in Fig.2 where the statistics of the final state pions is a crucial factor, there is no difference between with and without such statistics when we discuss density effect at low temperature (see Fig.2) which corresponds to the region of the first-order chiral phase transition in the $T-n_b$ plane. Instead of the statistics, the chiral restoration dominates the behavior of $g_{\sigma\pi\pi}$ and $ \Gamma_{\sigma\rightarrow 2\pi}$.
At $T=0$, the mixing phase of the first-order chiral transition is in the density region $0.33< n_b/n_0 < 1.56$ where $n_0$ is the baryon density in normal nuclear matter. At the up limit $n_b/n_0 = 1.56,\ m_\sigma < 2 m_\pi$, there is no $\sigma$ decay, while at the low limit $n_b/n_0 = 0.33,\ m_\sigma > 2 m_\pi$, $\sigma$ decay begins with maximum rate. With the decrease of the density, the decay rate and the coupling strength drop down from the maximums to their vacuum values, respectively. The fact that the maximum decay rate and the maximum coupling strength local at the low density limit of the mixing phase is directly due to the first-order chiral transition. The situation for $T=0.02\ GeV$ is similar, but the mixing region becomes narrower, $0.59<n_b/n_0<1.44$, and the jump of the decay rate (coupling strength) becomes smaller. As the temperature increases further, the density interval of the mixing phase becomes narrower and narrower, and the jump becomes smaller and smaller. Finally when approaching the end-point $E$ of the first-order transition at $T\simeq 0.03\ GeV$, the interval and the jump disappear. The decay begins at the point $E$ and proceeds faster and faster with the decrease of the density. The decay rate goes up from zero at the phase transition point $E$ to the maximum at low density. While the coupling strength has still a jump at the point $E$, the decay rate is continuous.   

To see the parameter dependence of the above discussion, we considered another set of parameters, $G=5.46\ GeV^{-2}, \Lambda=0.632\ GeV$ and $m_0=0.0055\ GeV$. It makes shifts of the magnitudes of $g_{\sigma\pi\pi}$ and $\Gamma_{\sigma\rightarrow 2\pi}$, but does not change their shapes, especially the behavior around the threshold temperature or the chiral phase transition point. 

From the above investigation, we see that when $\sigma$ decay into $2\pi$ happens in low density or low temperature region, the maximum decay rate does not locate in the vacuum, but is very close to the threshold point or just at the first-order chiral transition point. Only when the decay trajectory passes through the region around the end-point $E$ of the phase transition, the position of the maximum decay rate is far from the point $E$ where the sigma has minimum mass and therefore can be numerously produced. According to this picture, when the system of a relativistic heavy ion collision evolves in the low density or low temperature region, the $\sigma$ decay begins with maximum rate immediately after or even simultaneously at its starting point of creation. Since the $\sigma$ decay happens already at the freeze-out, there is no obvious non-thermal enhancement of pions. However, when the freeze-out occurs around the tricritical point, the $\sigma$ decay happens far from the freeze-out, there is really a non-thermal enhancement of pions which can be considered as a direct signature of chiral phase transition. 
\\
{\bf acknowledgments}:\\
This work was supported in part by the Natural Science Foundation of China under grant numbers 19845001 and 19925519.

\newpage
{\bf Figure Captions}\\ \\ 

{\bf Fig.1}: 
The Feynman diagram for $\sigma$ decay into $2\pi$ to the lowest order in $1/N_c$ expansion. The solid lines denote quarks and antiquarks, the dashed lines denote mesons.\\

{\bf Fig.2}:
The total decay rate $\Gamma_{\sigma\rightarrow 2\pi}$ and the coupling strength $g_{\sigma\pi\pi}$ as functions of the temperature at $n_b = 0$. The thick and thin solid lines indicate the decay rates with and without consideration of Bose-Einstein statistics for the final state pions. \\

{\bf Fig.3}:
The total decay rate $\Gamma_{\sigma\rightarrow 2\pi}$ (solid lines) and the coupling strength $g_{\sigma\pi\pi}$ (dashed lines) as functions of the baryon density scaled by $n_0$ for three values of the temperature. The black circle and black lozenge indicate the starting- and end-point of the mixing phase at $T=0$ and $T=0.02\ GeV$, respectively.

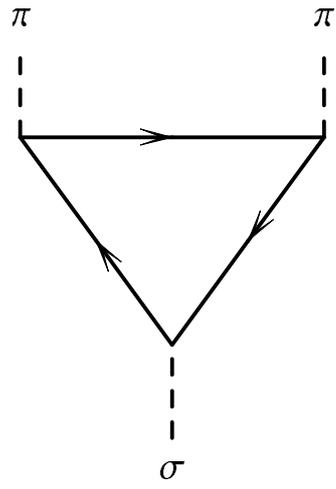
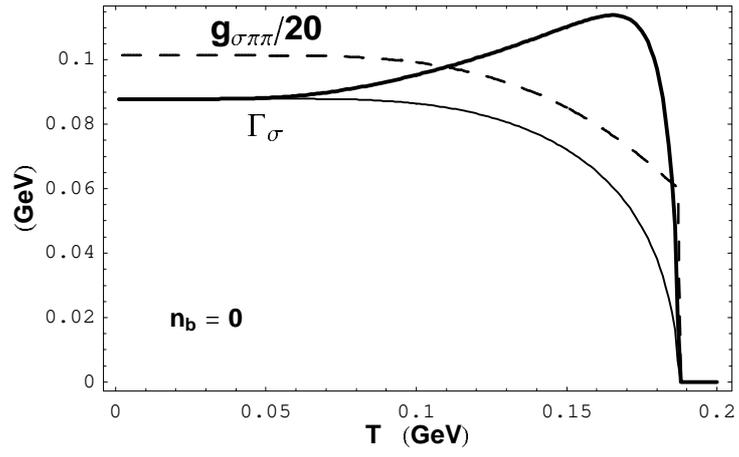
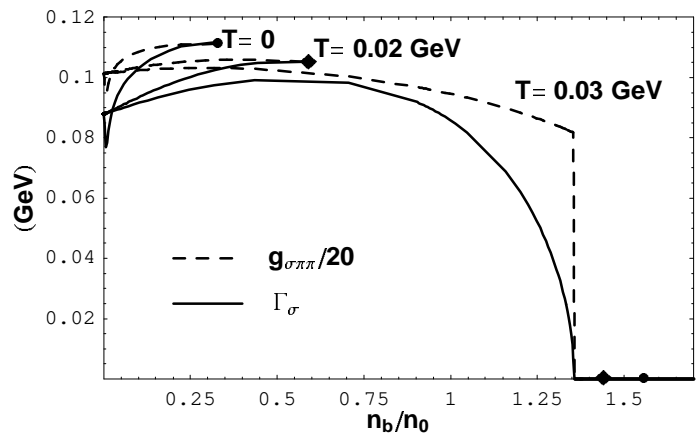
\end{document}